# Spreadsheet good practice:
# is there any such thing?

**David Colver**

Operis, 8 City Road,
London EC1Y 2AA, United Kingdom
dcolver@operis.com

**ABSTRACT**

Various techniques for developing spreadsheet models greatly improve the chance that the end result will not contain basic mechanical errors. However, for every discipline in which a given technique is useful, there is likely to be another in which the same technique works badly. As a result, the author urges that EuSpRIG does not succumb to internal or external pressures to champion a particular set of "best practices", because no such set is optimal in all spreadsheet applications.

1. APPROACH

In this paper, I am going to show some spreadsheet development ideas, which I find valuable. They are ideas that I believe in: they are methods that my company, Operis [Operis, 2010], uses when it develops financial models, and they are ideas that we teach to fee-paying customers when they come on the courses we provide. (Operis is by a long way the largest provider of financial modelling training, with over 10,000 man-days of courses attended in the last seven years.)

After I show you each idea, I am going to rubbish it.

My aim is to show you that what works well in one set of circumstances works badly in others. My goal is to discredit, once and for all, any temptation by insiders or calls from outsiders for EuSpRIG to set down some best practice guidelines for spreadsheets, despite the fact that such advice already exists [Read & Batson, 1999][Bewig, 2005][Raffensperger, 2001]. With very few exceptions, there is in my opinion no such thing as universal spreadsheet good practice. This opinion is supported by Grossman "Best practices are situation-dependent" [Grossman, 2002].

2. WHERE TO BEGIN

The first principle that Operis teaches is where to start when building a model. "If you fall asleep for the rest of the day, or if you don't absorb anything else while you are here, please try to remember the next sentence. Design the output first."

Designing the output means laying out the reports that your model is going to produce. In many financial models, this means setting out a cash flow, a profit and loss statement, a balance sheet, and a host of supporting schedules. But such models are by no means the only applications for spreadsheets, and users of spreadsheets in other disciplines will have different sorts of reports to produce.

These mocked up reports don't have any numbers in them at this stage. But they do have all the headings, neatly set out; and they do have the margins set properly, and the headers and footers, and are laid out so they fit on the page properly.

Setting these reports up is a good first step:

- Its work you are going to have to do eventually anyway; you might as well do it right away.
- It allows you to show your client, your boss and your colleagues what you intend to produce; if you have misunderstood the brief, they can probably tell you right away.
- It gives visible evidence of progress within a couple of hours of starting the project.
- If the client turns up unexpectedly, you can print out a neat version of the model in moments, and have something professional looking to show.

Most of all, the reports serve as an informal specification for the project. Completing the model becomes a simple matter of filling the lines in, one at a time, so the output serves as a route map for the assignment.

**Counterargument**

Designing the outputs first will indeed act as an informal specification for the project, but not all projects can live with such informality. Some projects are so important that they need quite formal specification to be negotiated with the project sponsors. In fact, a mock-up is not a specification as a traditional systems developer would have it. It is a non-functioning prototype, which is not the same thing at all.

Prototyping is itself a valuable tool in systems development, but banging together a mock up and seeing if the client likes the look of it is not the way we would build the systems that control a nuclear power station or fly an aeroplane. Where is the test plan? What are the acceptance criteria? How will we know that the spreadsheet gives the right answer when we see it? Many development methodologies teach that these are the areas to start with, not a rush to knock out a non-functional piece of Excel.

Even in circumstances where a formal specification is unnecessary, it is open to question whether designing the output first is a good thing to do. If your model produces standard financial statements, you can visualise what those will look like with a high degree of accuracy. But someone using a spreadsheet to develop some kind of mathematical or scientific theorem will probably go down many blind alleys, and be forced to tidy up his work so that it is in an acceptable state presentationally when he eventually does reach some kind of conclusion.

### 3. MODEL DESIGN: THE WORKINGS

The outputs we have mocked up don't do anything useful on their own because we haven't yet put any figures in them. By the time we have finished building a model, there will be figures, generated by spreadsheet formulae. But where should we put these spreadsheet formulae?

In deciding this, the principle we follow is that we wage war on long formulae. Any formula that is not so trivially short that it can be understood at a glance should be broken up and laid out over several lines. If you do that, though, to formulae on the output worksheets we have already prepared, you will tend to ruin their carefully prepared layout, and wind up with, a mess of unwanted intermediate workings.

The method that Operis teaches is that the calculations of a model should be on a separate workings page. That worksheet can take as many lines as it likes to calculate arbitrarily complicated results in easy steps. The conclusions from that work are then reported on the output sheet.

Besides providing the means to break up long calculations into short formulae without compromising the reports, separating the workings from the outputs has a number of practical advantages:

- Related items are calculated next to each other. The reporting of the expenditure on equipment, depreciation of that equipment and the resulting net book value of that equipment are reported on the cash flow, the profit and loss statement and the balance sheet, but the calculation of these intimately related items can be set out right next to each other. Formulae mention cells that are close at hand, which makes them easy to follow.

- If we placed the calculations on the reports, the formulae would have to operate in three dimensions. Besides being harder to read, such formulae are dramatically slower to calculate. On real world models, having the calculations on a single large workings sheet is about 5-10 times faster than dispersing them over many sheets.

It also in my view has a theoretical advantage: the way to calculate some results, and the way to display them, have nothing to do with each other. There is no other programming environment that I can think of that would contemplate mixing these items. Good system designs emphasise encapsulating design details, not jumbling them up.

Once we have our model written, we will start putting it to good use. One of the things we will want to explore is what happens if we take a different approach to the project being modelled. The way many people do this is to make copies of the workbook. And then overtype the numbers in it.

Quickly, you have many workbooks with many scenarios in them. Those scenarios are quite likely to exercise the model rather more vigorously than it has been exercised before, and expose various small bugs in it. Repairing them is difficult to do reliably, because you have many copies of the model that need identical repairs.

Our solution to this is to avoid having lots of models. We have just one model, with lots of scenarios in it. Each scenario has an input sheet of its own. We can switch between the scenarios very easily, by arranging for the workings sheet to take its data from one or other of the sheets. We normally provide a little menu option to do this, but in practice it activates some macro code of trivial proportions.

To sum up:

- we separate the inputs from the workings so that we can practice easy scenario management
- we separate the workings from the outputs to wage war on long formulae.

Figure 1 – Model Structure

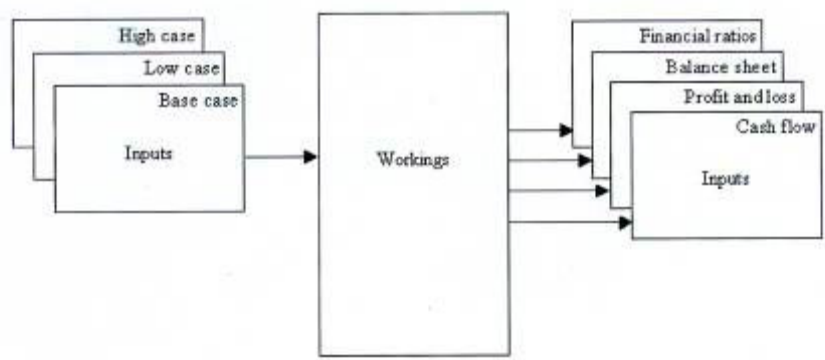

**Counterargument**

Separating the inputs, workings and outputs of a model has a nice, theoretical elegance to it, and it maps neatly on to how large scale data processing systems have worked for several decades. But spreadsheet models aren't large data processing systems: they are decision tools that work on devices that are not called personal computers for nothing.

By separating the inputs, workings and outputs you cause some lines in the model to appear three times over, once in each place. For someone who is not a proficient modeller, experienced spreadsheet user or even a fast typist, that is quite an overhead.

As well as making the model time consuming to develop, your method makes it harder to use. If someone doesn't like the figure reported by the model on one of the output sheets, they have to fathom out which inputs are responsible for it, and go and alter them appropriately. Those inputs are on a different worksheet, and quite hard to find. When naive spreadsheet users see a number they don't like, what they want to do is just type over it with a figure they prefer. With the inputs, workings and outputs all jumbled together, you have some chance of letting them do that. With these elements separated rigidly as you do, users of your model have to do something much less intuitive.

There is a particular circumstance where this issue really matters. A decade ago, models were used to structure a deal, and then thrown away when the paperwork was signed. Now the banks expect to see them updated every few months. Incorporating the historic actuals as they unfold is easy in a model that doesn't separate inputs from workings or workings from outputs: you just overtype formulae with real numbers. With the structure proposed in this paper, you are going to have to do something much more cumbersome to get historic values through the model from inputs to outputs.

There is one class of forecaster for whom this is a real pain. A broking analyst, paid to follow companies and express opinions about whether to buy or sell their shares, will have a spreadsheet for each firm that he is tracking. It will have historic figures, which then give way to projections. Every so often, perhaps once a quarter, the firm will announce some new results. He needs to update his spreadsheet with the new figures. In many cases, he will get two sets of figures: a press release with headline numbers, followed by the full accounts some weeks later.

For a while, therefore, his model has columns that are wholly historic; columns that are wholly projected; and one column that is a hybrid - see Figure 2. It has some actual numbers provided by the client, and others that have yet to be published and so must be derived. That's easy to do with a single worksheet in which inputs, workings and outputs are all the same thing; it is rather hard if the inputs, workings and outputs are rigidly separated.

Figure 2 – Historic, Hybrid & Projected Columns

| | Historic | | | Projected | | |
|---|---|---|---|---|---|---|
| | 2000 | 2001 | 2002 | 2003 | 2004 | 2005 |
| Sales | 98.4 | 101.2 | 103 | 110 | 121 | 132 |
| COGS | -54 | -60 | -62 | -70 | -75 | -80 |
| Admin costs | -10.1 | -11.2 | -12.5 | -13 | -14 | -15 |
| PBT | 34.3 | 52.4 | 28.5 | 27 | 32 | 37 |
| Tax | -13.72 | -20.96 | -11.4 | -10.8 | -12.8 | -14.8 |
| PAT | 20.58 | 31.44 | 17.1 | 16.2 | 19.2 | 22.2 |

The models Operis prepares concern themselves with large national infrastructure projects: roads, railways, hospitals, prisons. Financing them can take months, and during that time you do want to evaluate alternative scenarios. But the functionality of the model doesn't change much; all it is doing is applying the rules of double entry bookkeeping and calculating a few financial ratios. We can say that the logic of the model is stable, but the data is shifting.

Not all models are like that. A young associate in a bank's mergers and acquisitions department will often be asked to study the outcome of a merger between two companies. He will get the data from the companies' accounts. In the US, quite decent accounts appear every quarter, but in Europe, half-yearly accounts are more typical, and the mid-year report is very thin. The figures the banker has to work with therefore don't change at all for a whole year. With those figures he will prepare an analysis of what happens if company A buys company B. If that does not look sensible, he will wonder what happens if company B does a reverse takeover on company A; or if the two genuinely merge, or a new company is set up to buy the pair of them. Each of these transactions is fundamentally different and leaves him tearing his model down and rebuilding it.

In his case, it is the data of the model that is stable, but the logic that is shifting, quite the opposite situation from the Operis model. Separating the inputs from the workings to allow neat scenario management is a hindrance, not a help, in his circumstances.

## 4. AUDIT TESTS

Separating the workings from the outputs is a key tool in waging war on long formulae, but it does bring with it one notable drawback. It is possible to complete the most perfect of computations in the workings area, only to misreport it among the outputs. An example of such a mistake would be to leave a line out of the balance sheet.

My fix to this is to have a fourth kind of sheet, an audit sheet. Anything we can test on that, we do. There are obvious tests, like whether each subtotal adds up correctly, and-whether the balance sheet balances. And there are less obvious tests, such as an exact reconciliation of the total of the inputs sheet to the principal outputs.

A typical model shipped by Operis will have a 25% audit overhead, that is, an audit sheet that is a quarter of the size of the workings sheet. It's amazing how frequently the tests save our bacon. When asked to add some new capability to a model, we will do it diligently and methodically and with much skill and care. When we think we are done, we look at the audit sheet, and almost always find that there are three or four audit tests that are now failing. Without the audit sheet, we would have been more likely to ship the spreadsheet with those faults in it.

**Counterargument**

Putting copious self-testing into a financial model is such an obviously good thing that it is hard to imagine anyone not liking it. But there are people who object to them: they are professional spreadsheet auditors.

From his point of view, all the audit sheet does is introduce code to a spreadsheet that adds nothing to the outputs. It's code he has to make a choice about. Either, he can check that the tests do what they purport to do, which will take time and duplicate much of his own testing. Or, he can mention in his report that he is excluding them from his scope of work, which looks rather peculiar. In our experience, about half the model auditing practices implore us to remove the audit sheet from the versions of the model that they are asked to look at.

# 5. NAMES

We think of spreadsheet formulae as using a coordinate notation to involve other spreadsheet cells in a calculation. But spreadsheet formulae don't have to be expressed in terms of formulae; they can be expressed in terms of more meaningful names. It is entirely possible to build substantial models without a single coordinate being present in any of the formulae.

There are all sorts of practical advantages to using names.

- The names make the model easier to read. This is particularly true if your organisation follows a standard naming convention, as Operis does.
- The column matching is automatic. So long as you line the different worksheets up, so that column Z refers to the same time period throughout the model, then a whole class of error, the ones involving picking up figures from the wrong column, is eliminated.
- Formulae that refer to cells a considerable distance away, or on another worksheet, become much easier to read.

Names promote code reuse. An analyst developing a spreadsheet that concerns a project in Mozambique can save himself the bother of coding the tax regime in that country if he can lift it out of an existing model. A straight copy from one spreadsheet to another is most unlikely to work without adjustment if the formulae are expressed in terms of coordinates. It is unreasonable to expect the two spreadsheets to be coordinate-consistent; if they were, they would be the same spreadsheets, and there would be no point writing the second one. But it is quite reasonable, in an environment that uses names as a matter of course, to expect the two spreadsheets to follow a consistent naming convention; and if they do, the code fragment has every chance of working first time when it is copied.

The best modellers have libraries of code fragments waiting for reuse in this fashion, the most used of them having being subject to independent audit dozens of times as components of finished models. Often these libraries are supplemented by mechanical systems for coupling the fragments together into models.

There is also a theoretical appeal to using names, in my view. The solution to a problem has nothing to do with the way it happens to be laid out on a page. Einstein wrote E=mc2; he did not write that the quantity he has defined in the second paragraph of page two of his paper is equal to the one written about three lines above, multiplied by the square of the quantity he wrote about in another paper two months earlier. Nor should a modeller express his solution in terms that are dictated by how he happens to have laid it out on the page (and which will alter if a few lines are inserted or deleted for cosmetic reasons). This point is very similar to the justification offered for separating the workings of a model from the outputs: the calculation should not be influenced by how the results happen to be laid out.

**Counterargument**

Operis certainly makes use of names in its own models, and shows in its courses how to work with them, but it is much the least popular part of its training. There are many organisations which buy in to the Operis way of building a model in every respect save one: they don't bother with the names. Operis has lost some modelling assignments because the prospective client has preferred not to receive a model peppered with names. So much so, that the latest version of OAK,
the Operis Analysis Kit, a spreadsheet development and audit tool, actually includes a facility for reversing the naming process, restoring formulae to the coordinate representation.

Why are names so unpopular? There are some objections to them.

- They take time to define. You need to be certain of some payback before making that investment.
- The expression Sheet l! A1 refers unambiguously to a particular cell. The cell Revenue could refer to any number of cells on any of the worksheets. (Or it could refer to something else altogether.) How do we know that Revenue is correctly defined? Misdefined names give rise to a whole class of error that does not exist at all using the coordinate notation.
- What I am referring to here is functionality in Excel. It is not present in other spreadsheet packages, and supported only lightly in packages that purport to read data in the Excel file format. Using these features reduces portability to other platforms.
- Microsoft itself has done a number of things to undermine support for names in recent versions of Excel.

But I think the main reason that names are disliked is that people don't know how to manipulate them. They are not taught. It would only take half a day to learn; but if even a small number of consumers of a spreadsheet are prevented from altering the model by the use of a technique that is unfamiliar, the technique becomes less acceptable to the whole organisation in certain circumstances. For them, meaningful names have actually made the spreadsheet less useful than meaningless coordinates.

## 6. UNITS

One of the most basic errors that can be made in a spreadsheet is to combine inappropriately cells which have incompatible units. One cell may be in thousands of pounds, and another in millions: they can't be added together without adjustment. Some methodologies urge that every quantity is systematically labelled as to units.

My own approach is to urge that all spreadsheet quantities should be converted at the earliest available opportunity from whatever inputs they happen to be entered into base units. It doesn't matter what the base units are, as long as they are consistently applied throughout the model. If you are sure that quantities are in base units, you can combine them at will without worrying about multiplying or dividing by conversion factors.

My own choice of base units is invariably dollars, euros, pounds, tonnes, barrels (of oil) and so on. This gives the greatest flexibility, because the output can be shown in thousands, millions or billions merely by reformatting the spreadsheet, and so can be adapted to suit different audiences.

**Counterargument**

This flexibility is great for a professional spreadsheet programmer but having numbers display differently from the underlying value is thoroughly confusing for non-expert users of the spreadsheet and introduces a more dangerous possibility of error than it cures.

Even the expert can be confused when the data is exported. Copying to other Excel spreadsheets, or to, for example, Access, preserves the unformatted value. Copying to Word, or writing the data out as CSV or TXT files, gives the formatted version.

## 7. SPACES IN FORMULAE

Let me now mention one thing that Operis does not teach, but which was mentioned at EuSpRIG last year. We were urged to put spaces in our formulae to make them more readable.

**Counterargument**

If asked what operators are available in Excel, most people will correctly identify the mathematical operators, +, -, *, / and don't forget ^ and %.

Pushed a little harder, seasoned spreadsheet users will recall the comparison operators, >,<,= and their combinations, and the concatenation operator & for text.

But actually there are three other operators, colon, comma and space:

- The colon is the between operator: it constructs a range which includes all the cells between two other ranges.
- The comma is the union operator: it constructs a disjoint range which is composed of the components of two other ranges.
- The space is the intersection operator: it constructs a disjoint range which is composed of the components of two other ranges.

Space characters are therefore not white space in Excel. In early versions of Excel, you could only put a space in a formula if you meant it to be the intersection operator (or, obviously, as part of a literal, in quotes). Later versions relaxed the parser so that it would accept redundant spaces in the middle of formulae.

Nevertheless, I would no more put surplus space operators into my formulae than I would put surplus minus signs into them.

## 8. CONCLUSION

We have looked at issues ranging

- from the macro ...: which order to write a model, how to structure it;
- …through the middling ... : whether to use names or not, what units to choose;
- ... to the micro issue: whether or not it is helpful to add spaces for readability.

In every example, we can produce good, coherent arguments for following a particular policy, and find real world circumstances in which that policy would be quite the opposite of helpful.

At previous EuSpRIG meetings, I have been very uncomfortable when speakers have promoted their favourite way of doing things as the definitive method, good in all circumstances. Developing models is an exercise in engineering. It involves trade-offs. You can buy a fast car, or a cheap car, or one that carries lots of people. You may even find a valuable blend of two of these things. But you won't manage to get all three qualities in one car. Similarly, you can build a model that is laid out formally, or relatively easy to change, or very accessible to naive users; but you won't get all of them.

The methods I use are the ones that work for me, that make the trade-offs (see Figure 3) most helpful in my field of work. I know of other fields in which they will work very badly. There is a relatively high fixed cost to setting out a model in a structured fashion, and to using names.

- In our experience, the result is that the model scales better with complexity, but the spreadsheet needs to be long-lasting to get a useful payback (point A below).

- For those building models that are simpler, or short-lived, the fixed cost is not worth paying (point B below).

Figure 3 – Development cost / Model complexity trade off

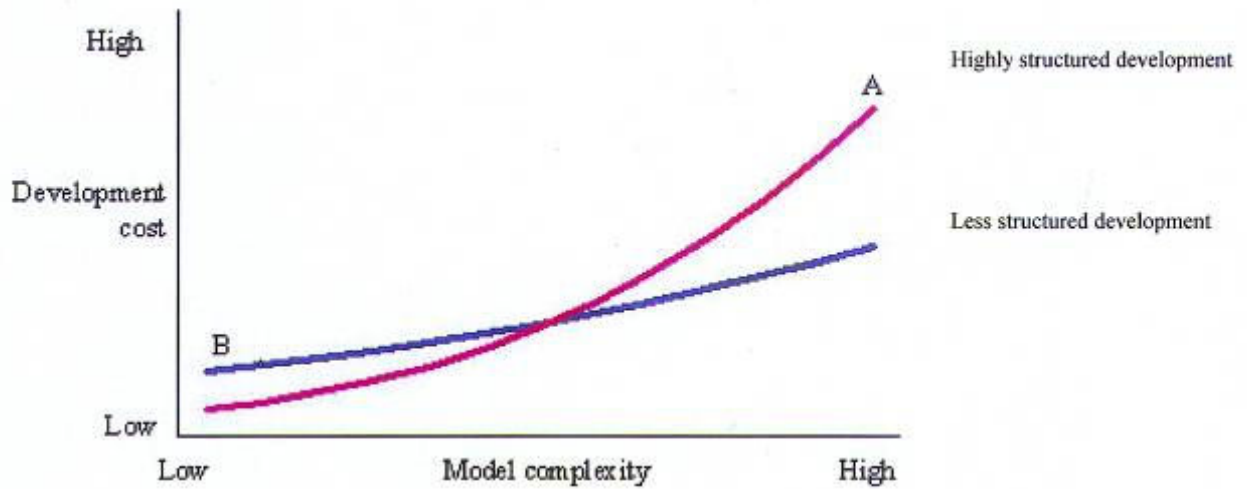

In closing, I urge EuSpRIG to be very thoughtful before it promotes any particular best practice. About the only techniques that I believe in as having a near universal application are:

- keep it simple: avoid overmodelling, or modelling beyond the resolution of your data
- have some idea of how you will know the finished article when you see it
- avoid long formulae
- don't use the OFFSET or INDIRECT functions
- introduce circular logic into a spreadsheet if you can explain the mathematical behaviour of the resulting function.